\def\year{2018}\relax
\documentclass[letterpaper]{article} 
\usepackage{aaai18}  
\usepackage{times}  
\usepackage{helvet}  
\usepackage{courier}  
\usepackage{url}  
\usepackage{graphicx}  
\usepackage{float}
\usepackage{bm}
\usepackage{amssymb}
\usepackage{amsmath}
\usepackage{amsthm}
\usepackage{array}
\newtheorem{myDef}{Definition}
\newtheorem{myTheo}{Theorem}
\newtheorem{myLemma}{Lemma}
\usepackage{multirow}
\usepackage{appendix}
\usepackage{bbm}
\usepackage{setspace}
\usepackage[linesnumbered,ruled]{algorithm2e}
\begin{document}
%
\title{BLAG: Bandit On Large Action Set Graph\vspace{-9mm}}\vspace{-5mm}
\author{Yucheng Lu, Xudong Wu, Jingfan Meng, Luoyi Fu, Xinbing Wang}
\maketitle
\begin{abstract}
\vspace{-1mm}
Information diffusion in social networks facilitates rapid and large-scale propagation of content. However, spontaneous diffusion behavior could also lead to the cascading of sensitive information, which is neglected in prior arts. In this paper, we present the first look into adaptive diffusion of sensitive information, which we aim to prevent from widely spreading without incurring much information loss. We undertake the investigation in networks with partially known topology, meaning that some users' ability of forwarding information is unknown. Formulating the problem into a bandit model, we propose \textbf{BLAG} (\textbf{B}andit on \textbf{L}arge \textbf{A}ction set \textbf{G}raph), which adaptively diffuses sensitive information towards users with weak forwarding ability that is learnt from tentative transmissions and corresponding feedbacks. \textbf{BLAG} enjoys a low complexity of $O(n)$, and is provably more efficient in the sense of half regret bound compared with prior learning method. Experiments on synthetic and three real datasets further demonstrate the superiority of \textbf{BLAG} in terms of adaptive diffusion of sensitive information over several baselines, with at least 40\% less information loss, at least 10 times of learning efficiency given limited learning rounds and significantly postponed cascading of sensitive information.
\end{abstract}
\vspace{-5mm}
\section{Introduction}
The prevalence of massive social medias facilitates information diffusion, which plays an important role in content sharing and spreading \cite{bakshy2012role}. While the low-cost diffusion can easily lead to large-scale propagation called information cascading, the unconstrained cascading behavior could meanwhile cause the sensitive information to be incautiously propagated over the network. Here the sensitive information can refer to any kind of information that needs to be prohibited from cascading such as rumors, personal information, trade secrets, and etc. Generally, sensitive information has two features: a) \textit{Hard to capture}, meaning that it is hard for classification because of being user-specific and hard for detection due to huge volume of daily information consumption. b) \textit{Time-sensitive}, meaning that it can quickly fade out with time elapse (e.g., rumors about US presidential election in 2016 is no longer sensitive in 2017).\\
\indent The two above features, consequently, render it rather difficult to prevent the sensitive information from wide spread by  targeting the sensitive information itself. However, it will be a lot easier if we instead target those individuals who may carry the sensitive information. In other words, we can adaptively adjust the diffusion of information based on users' attributes in hope of subsiding the cascading of sensitive information while preserving the propagation of non-sensitive one. Such adaptive transmission in the sense of sensitive information has unfortunately received no prior attention amongst the intensively studied topic of information diffusion \cite{guille2012predictive}\cite{wang2012diffusive}\cite{yang2010modeling}\cite{yang2015rain}. The only two recent attempts that share the closet correlation with the issue belong to \cite{xu2015modeling} and \cite{giakkoupis2015privacy}. Particularly, considering the time sensitiveness of information, \cite{xu2015modeling} points out that decreasing users' information forwarding probability with time results in limited size of cascading. However, the method may blindly block a lot of non-sensitive information, thus leading to severe information loss.  \cite{giakkoupis2015privacy} aims not to reveal much about the users' own opinions on items by reposting them based on users' opinions and followers. The idea nevertheless requires that users have a full knowledge of network topology, which violates real practices. \\
\indent Regarding both concerns, \textbf{we are motivated to undertake the first investigation of adaptive diffusion of sensitive information with partially known network topology}, which, in our context, refers to that some users have no prior knowledge of their neighbors. Assuming that users have different levels of forwarding ability that are positively related to their degrees, we intend to adaptively balance their outward transmission behaviors by constraining the sensitive information transmission from those with strong forwarding ability while promoting more transmissions from users with weak forwarding ability. The idea of guiding the information diffusion towards users with weak forwarding ability can potentially incur less information loss than simply limiting the diffusion. Since some users' forwarding ability is unknown due to lack of prior knowledge of their neighbors, adaption should be based on tentative tranmissions and corresponding feedbacks. \\
\indent  Technically, we convert the above adaptive diffusion into a \textbf{Constrained Combinatorial Multi-Arm Bandit} (\textbf{CCMAB}) model, where we propose a learning algorithm \textbf{BLAG} (\textbf{B}andit on \textbf{L}arge \textbf{A}ction set \textbf{G}raph). While we defer more details of \textbf{BLAG} later, here we briefly delineate its main ingredients of: \textbf{BLAG} makes transmission policy based on current estimation of users' forwarding ability and update the policy based on the new feedback, which is calculated by the expected number of nodes receiving information under adopted policy with observation noise. The update of policy balances the trade-off between trying to get the optimal policy and minimizing the feedback in the next trial. \textbf{BLAG} turns out to enjoy a low complexity of $O(n)$, and provably returns half of the time-limited regret bound compared to previously proposed \textbf{CUCB}(\textbf{C}ombinatorial \textbf{U}pper \textbf{C}onfidence \textbf{B}ound) strategies \cite{gai2010combinatorial}\cite{chen2013combinatorial}.\\
\indent Our key contributions can be summarized as follows:
\begin{itemize}
	\item We take the first look into adaptive diffusion of sensitive information in networks with partially known topology, and formulate it into a novel constrained combinatorial bandit problem.
	\item We propose a learning algorithm \textbf{BLAG} that involves bandit on large action set for adaptive diffusion. \textbf{BLAG} theoretically reduces the time-limited regret bound by half compared to classic \textbf{CUCB} bandit algorithms.
	\item We perform extensive experiments on both synthetic and real data sets, which confirm the superiority of \textbf{BLAG} over several baselines in terms of at least 40\% less information loss, at least 10 times higher learning efficiency as well as postponed cascading of sensitive information.
\end{itemize}
\vspace{-4mm}
\section{Preliminary}\vspace{-1mm}
\subsection{Network Model}
We model the social network as an undirected graph $G=(V,E)$ with neither self-loops nor multiple edges between any two nodes. Each node is classified as either a \textbf{sensitive} node or a \textbf{non-sensitive} one. In correspondence to real social networks, sensitive nodes can refer to the individuals who hold plenty of sensitive information. A strict criterion for classifying nodes into sensitive and non-sensitive ones is not our focus and will not affect our later analysis. In each \emph{time slot}, sensitive information can only be transmitted from a \textbf{sensitive} node to a \textbf{non-sensitive} one, and there can be \underline{multiple} transmissions within one slot. In addition, the transmission of non-sensitive information also occurs between nodes. Here our focus falls primarily on extracting and analyzing the transmission of sensitive information, which, different from non-sensitive information, would be limited by the desired adaptive diffusion that we will introduce shortly. Each edge $e\in E$ has a weight representing the transmission probability of sensitive information  between the two nodes it connects. We assume that the weight follows a uniform distribution $U(0,\xi)$, where $\xi$ is a sufficiently small number, implicitly implying that sensitive information is hard to capture.
 A non-sensitive node may turn into a sensitive one in a
\emph{time slot} as long as it receives enough amount\footnote{The exact amount is is beyond the focus of this paper. In the
actual social network, only a user receiving tons of sensitive
information should be monitored.} of sensitive
information. Let $G'$ denote the subgraph induced by sensitive nodes and $G\backslash G'$ the subgraph induced by non-sensitive nodes.

 As mentioned earlier, we consider the network with partially known topology, which, in our setting, is associated with whether non-sensitive nodes are \emph{informed} or \emph{uninformed}, defined as follows:
\begin{myDef}
	Informed nodes refer to the non-sensitive nodes with known one-hop neighbors, while uninformed nodes are non-sensitive nodes with unknown one-hop neighbors.
\end{myDef}
Based on Definition 1, we further give the definition of partially known topology, which we call \textbf{semi-informed network} throughout the paper:
\begin{myDef}
	A semi-informed network is a network where both informed and uninformed non-sensitive nodes coexist along with sensitive nodes.
\end{myDef}

\indent For a better understanding, Fig. 1 further illustrates an example of information diffusion in a semi-informed network.

\begin{figure}[!h]
	\vspace{-9mm}\centering
	\includegraphics[width=0.48\textwidth]{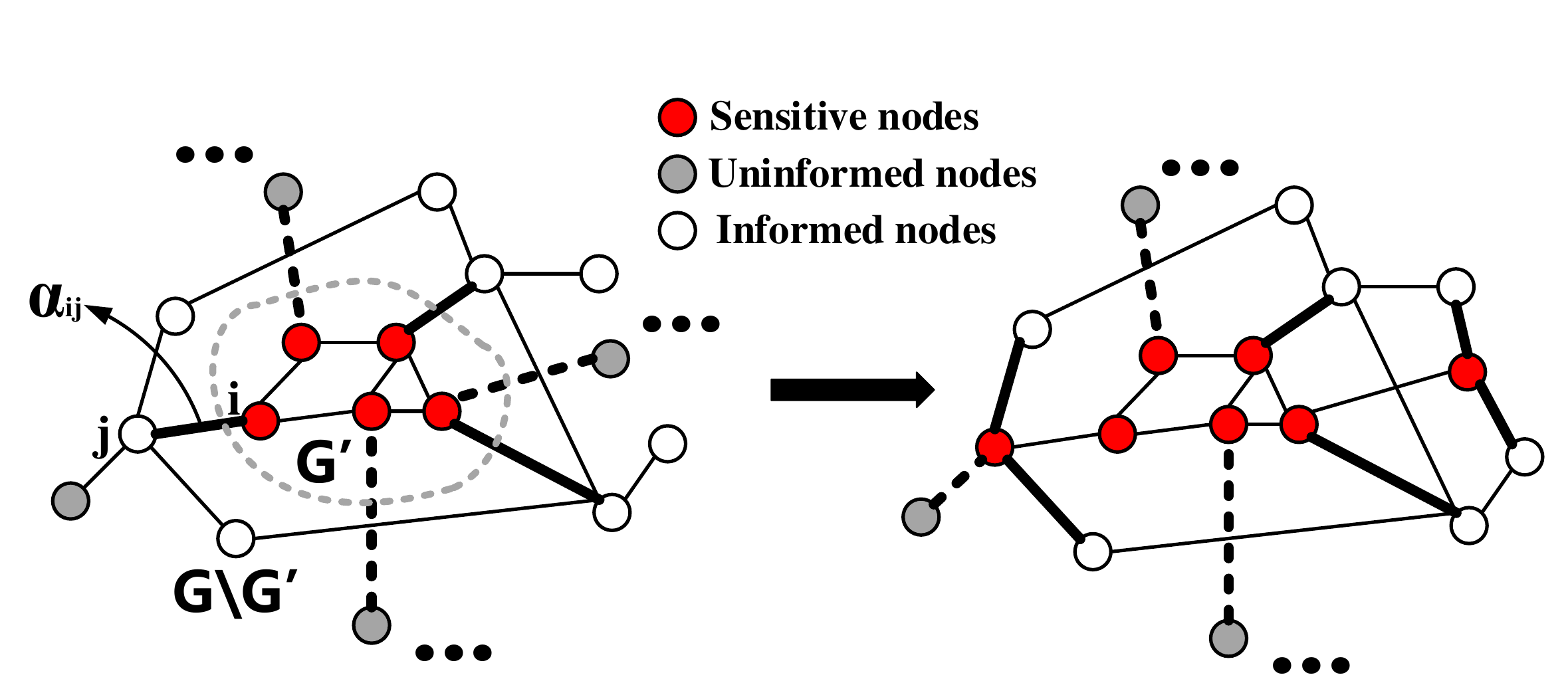}
	\vspace{-8mm}\caption{\small An illustration of a semi-informed network. The red nodes represent sensitive nodes connected by two types of non-sensitive nodes, i.e., informed nodes (white) and uninformed ones (grey). Information transmission occurs between nodes as long as there is an undirected path in between, and their sensitivity states subject to the foregoing conditions. Here node $j$ turns from a non-sensitive node into a sensitive one after receiving sufficient information from node $i$. $\alpha_{ij}$ is the probability $i$ sending sensitive information to $j$.
}
\end{figure}
\vspace{-6mm}
\subsection{Problem Statement}
Let $\mathfrak{E}$ denote the set of edges connecting nodes from $G'$ and uninformed nodes from $G\backslash G'$. The problem of interest in semi-informed network is to adaptively diffuse information among nodes by adjusting parameter on edges in $\mathfrak{E}$ based on destination's forwarding ability. In the present work, we simply associate a node's forwarding ability with its degree, with a high degree implying a strong forwarding ability. Our target is to adaptively diffuse sensitive information towards low degree nodes, while avoiding high degree nodes.\\
\indent As the degrees of the uninformed nodes are unknown, our target, equivalently, is to make \emph{trials} to these uninformed nodes and adaptively diffuse information based on the \emph{feedback}. We denote edges in $\mathfrak{E}$ with uninformed nodes being destination as \emph{target edges} and these connected uninformed nodes as \emph{target nodes}. In a time slot, let the $m$-dimensional vector $\overrightarrow{D}$ denote the degree set of $m$ \emph{target nodes}, with $\overrightarrow{D}(i)$ representing the degree of \emph{target node} $i$. As fomulated earlier, exact values of elements in $\overrightarrow{D}(i)$ are unknown. The $m$-dimensional vector $\overrightarrow{\beta_{0}}$ denotes original transmission probability on \emph{target edges}, with $\overrightarrow{\beta_0}(i)$ representing original transmission probability on edge $i$. Let $\overrightarrow{\Delta\beta}$ denote the variation of probability on \emph{target edges}, with $\overrightarrow{\Delta\beta}(i)$ represents variation of probability on edge $i$. Thus, our target above is to learn an optimal $\overrightarrow{\Delta\beta}^*$ within the time slot that minimizes
\begin{displaymath}
	\overrightarrow{D}\cdot (\overrightarrow{\beta_{0}}+\overrightarrow{\Delta\beta}^*)
\end{displaymath}
\indent This process, as we will demonstrate in sequel, is similar to a bandit problem, which technically interprets the adaptive diffusion problem in semi-informed networks.
\vspace{-2mm}
\section{Problem Formulation}
\subsection{Mapping Diffusion Problem into Bandit Model}
 We assume there are multiple but limited rounds of trials in one time slot. During the trials, we try to minimize $\overrightarrow{D}\cdot (\overrightarrow{\beta_{0}}+\overrightarrow{\Delta\beta}^t)$ where $\overrightarrow{\Delta\beta}^t$ represents variation vector in round $t$. We adapt information transmission probability to those uninformed nodes based on trials and feedbacks. To this end, we map the problem into a \textbf{Constrained Combinatorial Multi-Arm Bandit (CCMAB)} model. Particularly, we summarize below the mapping between \textbf{CCMAB} components and the key elements in the formed problem:
\begin{itemize}
	\item \vspace{-2mm}\underline{Trial:} A trial in round $t$ refers to a transmission policy determined by $\overrightarrow{\Delta\beta}^t$. The total number of trials between two time-slots is determined. Besides, we assume that the statuses of both sensitive and non-sensitive nodes remain unchanged in a time slot.
	\item \vspace{-2mm}\underline{Base-action:} In a combinatorial bandit model, a base-action \cite{chen2013combinatorial} is an atomic action in the action set. In our problem, base-actions are vectors with pair-wise non-zero elements. In adaptive diffusion, to maintain the overall probability on \emph{target edges}, for any probability decreasing on one edge, there must be a corresponding increasing probability of same amount on another edge. Based on this fact, there are pair-wise edges containing probability variation with zero summation. We regard these vectors representing pair-wise variation of edges as base-actions $\overrightarrow{\beta_i}$, each of which has the following characteristics:
	\begin{itemize}
		\item $\overrightarrow{\beta_i}$ only has two non-zero elements.
		\item $\sum_{j=1}^{m}\overrightarrow{\beta_i}(j)=0$.
		\item $\overrightarrow{\beta_i}(j) \in [-1,1], 1\leq j\leq m$.
	\end{itemize}
	\item \vspace{-2mm}\underline{Super-action:} A super-action \cite{chen2013combinatorial} is combination of several base-actions. However, in our problem, arbitrary combinations may not be valid. To determine whether two arms can combine in valid, we give \textbf{Definition 3}. It not only works for base-actions but for any base-action or super-action.
	\begin{myDef}
		Combination of any two arms $\overrightarrow{\beta_{1}}$ and $\overrightarrow{\beta_{2}}$ is valid if and only if $\forall i, 1\leq i \leq m$,  $0\leq \overrightarrow{\beta_{0}}(i) + \overrightarrow{\beta_{1}}(i) + \overrightarrow{\beta_{2}}(i)\leq 1$
	\end{myDef}
	The restriction holds due to the natural bound of probability. An illustrative example of valid combination of two base actions is demonstrated in Fig. 2.
	\begin{figure}[!h]
		\vspace{-4mm}\centering
		\includegraphics[width=0.48\textwidth]{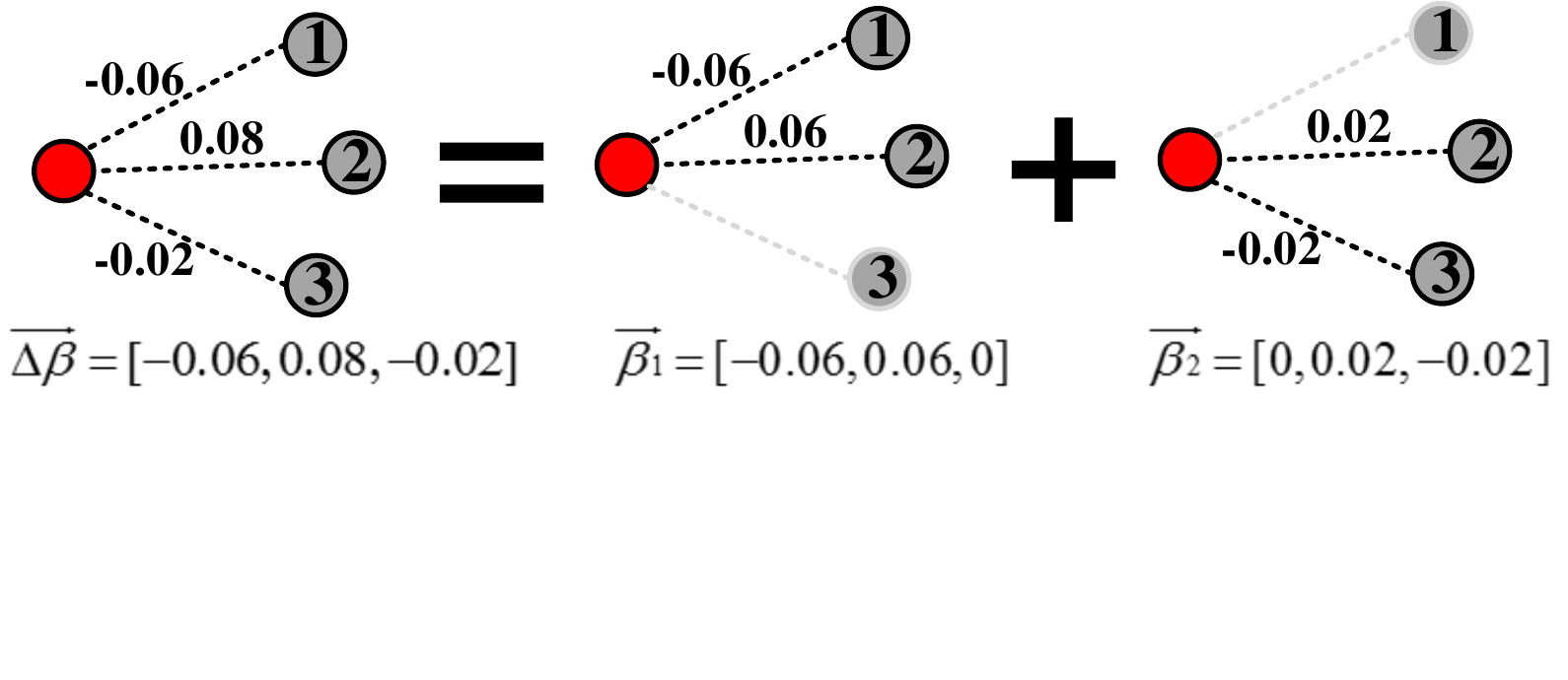}
		\vspace{-22mm}\caption{\small A toy example of combining base-actions $\overrightarrow{\beta_1}$ and $\overrightarrow{\beta_2}$ into a super-action $\overrightarrow{\Delta\beta}$.}
	\end{figure}
	\vspace{-3mm}
	\item \underline{Reward:} In our problem, reward refers to feedback of an adopted super-action, in the form of $\overrightarrow{D}\cdot\overrightarrow{\Delta\beta}+\sigma$, where $\sigma$ is the noise in observation subjected to Gaussian distribution. We assume reward of base-actions combined in the observed super-action can be observed individually. Other than that, we assume that each base-action contains an unreliable estimation before the first trial. (In the real world, it could be context or estimation of a user's forwarding ability)
\end{itemize}
\vspace{-4.5mm}
\subsection{Learning Process}
\vspace{-1mm}
Upon the mapping from the diffusion problem into the bandit model, the main focus in learning process is to get the minimum $\overrightarrow{D}\cdot (\overrightarrow{\beta_{0}}+\overrightarrow{\Delta\beta}^t)$ from super-action $\overrightarrow{\Delta\beta}^t$ selected in round $t$. This target is equivalent to minimizing the regret, which means the variation of the reward between the selected super-action and the optimal super-action. Since a lower regret demonstrates better performance of a bandit policy, the learning process is to obtain as low time-limited regret bound as possible within the total rounds in a slot.
\vspace{-2mm}
\section{Algorithm}
\vspace{-1mm}
The mapping and learning that we have just mentioned may provoke two major issues. One is the huge size of base-action set. Based on our assumption, the size of the base-action set could reach $O(m^2)$ (The number of vectors with pair-wise elements is $\binom{m}{2}$ and each pair could have multiple variations). Another issue is the limited number of trials in a slot. Here we assume that the size of the base-action set is too large to be trialed all over in one slot.

To tackle the large base-action set problem, we map base-actions into a weighted graph and denote the graph as \textbf{Action Set Graph} ($ASG$). As for the limited learning rounds problem, we let our algorithm globally follow $\varepsilon-greedy$ process. Based on techniques, we propose a learning method called \textbf{BLAG} (\textbf{B}andit on \textbf{L}arge \textbf{A}ction set \textbf{G}raph).
\vspace{-1mm}
\subsection{Action Set Graph}
We start with the action set graph, which represents whether combination of two \underline{base-actions} is valid. The topology of $ASG$ follows two rules: (1) Each node in $ASG$ represents a base-action $\overrightarrow{\beta_{i}}$, and (2) the weight of each nodes represents the current estimated reward of the corresponding base-action. If the combination of the two base-actions is valid, their corresponding nodes in $ASG$ are connected by an unweighted edge. Obviously, any valid combination is a clique in $ASG$, and nodes in a clique are pair-wise connected.
\vspace{-2mm}
\subsection{Limited Learning Rounds}
The key point in the learning process is to balance the \emph{exploration-exploitation trade-off} while considering limited learning rounds. Two widely adopted methods of tackling the issue belong to $\varepsilon-greedy$ and $UCB$ (Upper Confidence Bound)\cite{auer2002finite}-based approaches. In our algorithm, we adopt $\varepsilon-greedy$ with $\varepsilon$ decreasing with time. As we will theoretically demonstrate later, such strategy is provably twice efficient than the alternatives in the learning process. Here we summarize the two key procedures, i.e., exploration and exploitation, as follows:
\begin{itemize}
	\item \textbf{Exploration:} The target in this procedure is performing trial to get as much context of the model  as possible. Since each base-action can only be observed once in each round, our strategy here is to select as many base-actions as possible.
	\item \textbf{Exploitation:} The target in this procedure is to get minimum reward under current context. That is, it selects the combination with the minimum summation of reward.
\end{itemize}
\vspace{-4mm}
\subsection{Exact Bandit Algorithm}
\indent Now we introduce our designed algorithm \textbf{BLAG}, the global process of which follows an $\varepsilon-greedy$ process.That is, at the beginning of each round, we decide exploration or exploitation by a given time-decreasing $\varepsilon$. Later we will show that our algorithm performs better than Combinatorial UCB algorithm when the number of learning rounds is limited.\\
\indent Generally, \textbf{BLAG} can be decomposed into three parts: \textbf{Exploration}, \textbf{Exploitation} and \textbf{Update parameters}.\\
\indent As noted earlier, the target in exploration procedure is to get a large size combination. Regarding $ASG$, this can be treated as a Maximum Clique like problem in the graph theory. However, the issue is further complicated in our scenario since the algorithms proposed in the Maximum Clique problems cannot guarantee the valid combination expected in our problem. We thus tackle the issue by designing a $BFS$-like procedure in our algorithm. That is, we randomly choose a node in $ASG$ and iterate its neighbors, if the neighbor is not conflicting with the current combination, we add it to the combination. A difference between $BFS$ and our procedure here is we only iterate the one-hop neighbor of the chosen node. Because any multi-hop node is conflicting with the chosen node, thus it must not be chosen. The pseudo code of exploration procedure is given in \textbf{Algorithm 1}.\\
\begin{algorithm}[!h]\footnotesize
	\caption{Exploration procedure}
			\begin{spacing}{0.65}
	\KwIn{$ASG$}
	\KwOut{A super-action}
		$u\leftarrow$RANDOM($ASG,1$)\; \tcc*[f]{RANDOM($S, n$) returns $n$ random nodes in graph $S$.}\;
		$combination\leftarrow\{u\}$\;
		$iteration\leftarrow 1$\;
		\For{$v$ in $\Gamma(u)$}
		{
			$iteration\leftarrow iteration+1$\;
			\If {$iteration>m$}
			{
				Break\;
			}
			\tcc*[f]{VALID($a_1,a_2$) returns a boolean value of whether the combination of vectors $a_1$ and $a_2$ is valid or not.}\;
			\If {VALID($combination,v$)}
			{
				$combination\leftarrow combination\cup\{v\}$\;
			}
		}
		return $combination$\;
\end{spacing}
\end{algorithm}
\indent Comparatively, the target in exploitation procedure is to select combination with minimum cumulative reward. Regarding $ASG$, this is a Maximum Weighted Clique like problem in the graph theory. Similar to the exploration procedure, algorithms proposed in the Maximum Weighted Clique problems cannot guarantee valid combination here. In our algorithm, we use a greedy strategy, which is to say, we iterate nodes in $ASG$ based on their estimated reward. If a node is not conflicting with current combination and can help us get a smaller reward summation, we add it to the combination. \textbf{Algorithm 2} provides the pseudo code of the exploitation procedure.\\
\begin{algorithm}[!h]\small
	\caption{Exploitation procedure}
\begin{spacing}{0.65}
	\KwIn{$ASG$}
	\KwOut{A super-action}
		$ActionPool\leftarrow$RANDOM($ASG,\lfloor\sqrt{m}\rfloor$)\;
		$combination\leftarrow\varnothing$\;
		\While {$ActionPool\neq \varnothing$}
		{
			$v=$MIN($ActionPool$)\;
			\tcc*[f]{MIN($S$) returns item with smallest value in set $S$.}\;
			\If {$\mu_{v,t}>0$}
			{
				Continue\;
			}
			$ActionPool\leftarrow ActionPool\backslash \{v\}$\;
			\If {VALID($combination,v$)}
			{
				$combination\leftarrow combination\cup\{v\}$\;
			}
		}
		return $combination$\;
\end{spacing}
\end{algorithm}
\indent Combined with the updating procedure, the whole pseudo code of \textbf{BLAG} is shown in \textbf{Algorithm 3}. After jumping out of procedure of forming combination, \textbf{BLAG} has a final procedure of updating estimation of reward and selected time of each nodes selected.
\vspace{-4mm}
\begin{algorithm}[!h]\small
	\caption{BLAG}
\begin{spacing}{0.65}
	\KwIn{$ASG$, initialized $\varepsilon_0$, learning round $T$}
	\KwOut{a sequence of super-actions}
		\For{$t=1$ to $T$}
		{
			$\varepsilon_t \leftarrow \frac{\varepsilon_0}{\sqrt{t}}$\;
			\eIf {$\varepsilon_t$}
			{
				$combination\leftarrow$Exploration($ASG$)\;
			}
			{
				$combination\leftarrow$Exploitation($ASG$)\;
			}
			\For{$i$ in $combination$}
			{
				$\mu_{i,t}\leftarrow[(t-1)*\mu_{i,t-1}+reward(i)]/t$\;
				$T_{i,t}\leftarrow T_{i,t-1}+1$\;
			}
		}
\end{spacing}
\end{algorithm}
\vspace{-5mm}
\subsection{Complexity}
	In the exploration procedure, we jump out of iteration when the iteration times reaches $m$. Thus in this procedure the complexity is $O(m)$. In the exploitation procedure, we iterate $\lfloor \sqrt{m}\rfloor$ nodes in $ASG$ and \textbf{MIN} contains a sorting sub-procedure, which brings $O(m)$ complexity. In the last step, we iterate the selected nodes, which is upper bounded by $m$, also brings $O(m)$ complexity. As a result, the overall complexity of our algorithm is $O(m)$.
\section{Regret Analysis\footnote{For page restrictions, the detailed deduction of Theorem 1, Lemma 1, Eqs. (1), (2) and (3) are available at https://github.com/EugeneLYC/AAAI18.}}
We proceed to analyze the regret bound brought by \textbf{BLAG}, and meanwhile provably show \textbf{BLAG} exhibits half expected regret than previous approach.
\vspace{-2mm}
\subsection{Time-Limited Regret Bound of BLAG}
\indent In $\varepsilon-greedy$, the total reward in round $t$ can be written as:
\begin{displaymath}
\small
	\mathbb{E}[\overrightarrow{D}\cdot\overrightarrow{\Delta\beta}^t] = \varepsilon\mathbb{E}[\overrightarrow{D}\cdot\overrightarrow{\Delta\beta}_{ep}^t] + (1-\varepsilon)\mathbb{E}[\overrightarrow{D}\cdot\overrightarrow{\Delta\beta}_{et}^t]
\end{displaymath}
where ${\Delta\beta}_{ep}^t$ is the super-action adopted from exploration and ${\Delta\beta}_{et}^t$ is the super-action adopted from exploitation. Let $\mu_{i,t}$ be estimated reward of base-action $i$ at round $t$, then we have
\begin{displaymath}
\small
	\mathcal{P} \left( |\overrightarrow{D}\cdot\overrightarrow{\beta}_i - \mu_{i,t} | > \delta  \right) \leq 2e^{-\frac{T_{i,t}\delta^2}{2\sigma^2}}
\end{displaymath}
where $T_{i,t}$ is the time base-action $i$ has been selected by round $t$. The base-actions choosen in exploitation satisfy satisfies: $\mathbb{E}\left[\sum_{i\in\mathcal{S}_t } \mu_{i,t}\right] \leq \alpha \mathbb{E}\left[\sum_{i\in\mathcal{S}^*_t } \mu_{i,t}\right]$, where $\alpha$ is the approximation factor of \textbf{Algorithm 2}, $\mathcal{S}_t$ is the output combination of \textbf{Algorithm 2}, and $\mathcal{S}^*_t$ is the set of base-actions in the optimal super-action. Without loss of generality, we sort $\overrightarrow{D}$ and let $B_0\triangleq\sum_{j=1}^{m} \overrightarrow{\beta}_{0}(j)$, $B^*\triangleq\sum_{j=1}^{B_0} \overrightarrow{D}(j) - \overrightarrow{D}\cdot\overrightarrow{\beta}_0$ and $B^\times\triangleq \sum_{j=m-B_0+1}^{m} \overrightarrow{D}(j) -\sum_{j=1}^{B_0} \overrightarrow{D}(j) $.
\begin{myTheo}
For any valid combination $\overrightarrow{\beta}$, $\sum_{j=1}^{m} \overrightarrow{\beta}(j) = B_0$: $\overrightarrow{D}\cdot\overrightarrow{\beta}- \overrightarrow{D}\cdot\overrightarrow{\beta}_0 \geq B^*$
\end{myTheo}
\vspace{-3mm}
\begin{myLemma}
For any two valid combinations $\overrightarrow{\beta}_1,\overrightarrow{\beta}_2$: $\overrightarrow{D}\cdot\overrightarrow{\beta}_1- \overrightarrow{D}\cdot\overrightarrow{\beta}_2\leq B^\times$.
\end{myLemma}
Let event $\mathcal{F}_t$ be $\left\{ |\overrightarrow{D}\cdot\overrightarrow{\beta}_i - \mu_{i,t} |\leq \frac{c\sigma}{\sqrt{T_{i,t}}} , \quad \forall i \in \mathcal{V}\right\}$, where $\mathcal{V}$ is the base-action set and $c$ is a contant. Based on Theorem 1 and Lemma 1, we can get
\begin{equation}
\small
	\begin{aligned}
	\mathop{\mathbb{E}}[\overrightarrow{D}\cdot\overrightarrow{\Delta\beta}_{et}^t-\alpha\overrightarrow{D}\cdot\overrightarrow{\Delta\beta}^*]\leq c\sigma\sum_{i\in\mathcal{S}_t} \frac{1}{\sqrt{T_{i,t}}}+ B^\times\mathcal{P}(\overline{\mathcal{F}_t})
	\end{aligned}
\end{equation}
The only undetermined element in $\mathop{\mathbb{E}}_{\mathcal{F}_t}\left[ \sum_{i\in\mathcal{S}^*_t }\mu_{i,t}\right]$ is the observation noise. Let  ${\mathcal{S}^*_t}'$ be the set of base-actions in optimal super-action. From $\sum_{i\in{\mathcal{S}^*_t}' }\mu_{i,t} \geq \sum_{i\in\mathcal{S}^*_t }\mu_{i,t}$, we have $\overrightarrow{D}\cdot\overrightarrow{\Delta\beta}^* = \mathop{\mathbb{E}}_{\mathcal{F}_t}\left[ \sum_{i\in{\mathcal{S}^*_t}' }\mu_{i,t}\right] \geq \mathop{\mathbb{E}}_{\mathcal{F}_t}\left[ \sum_{i\in\mathcal{S}^*_t }\mu_{i,t}\right]$. Now, we can calculate regret bound from exploitation in $T$ rounds:
\begin{equation}
\small
	\begin{aligned}
	\mathbb{E}\left[\sum_{t=1}\cdot\overrightarrow{D}\cdot\overrightarrow{\Delta\beta}_{et}^t-\alpha\overrightarrow{D}\cdot\overrightarrow{\Delta\beta}^*\right]\leq 2c\sigma  M\sqrt{T}+ B^\times\sum_{t=1}\cdot\mathcal{P}(\overline{\mathcal{F}_t})
	\end{aligned}
\end{equation}
Let $c\geq \sqrt{2\ln 2B^\times MT}$, thus we can get
\begin{displaymath}
\small
B^\times\sum_{t=1}\cdot\mathcal{P}(\overline{\mathcal{F}_t}) \leq B^\times MT\cdot 2e^{-\frac{c^2}{2}} \leq 1
\end{displaymath}
Thus this term can be ignored. Let $\varepsilon_t = \frac{\varepsilon_0}{\sqrt{t}} $, and we can finally get the regret bound of \textbf{BLAG}:
\begin{displaymath}
\small
\begin{split}
&\mathbb{E}[R_{\textbf{BLAG}}] =\\& \mathbb{E}\left[\sum_{t=1}\cdot\overrightarrow{D}\cdot\overrightarrow{\Delta\beta}_{et}^t-  \alpha T\overrightarrow{D}\cdot\overrightarrow{\Delta\beta}^*+ \sum_{t=1}\cdot\varepsilon_t \left(\overrightarrow{D}\cdot\overrightarrow{\Delta\beta}_{ep}^t-\overrightarrow{D}\cdot\overrightarrow{\Delta\beta}_{et}^t\right)\right] \\&\leq 2c\sigma  M\sqrt{T}+ B^\times \sum_{t=1}\cdot\varepsilon_t +1 \leq 2c\sigma  M\sqrt{T} + 2B^\times \sqrt{T}+1
\end{split}	
\end{displaymath}
The last step holds because $m$ is a sufficiently large number.
\vspace{-6mm}
\subsection{Superiority of the Derived Regret}
We further prove \textbf{BLAG} has superior performance on regret bound to the alternative algorithm \textbf{CUCB} (\textbf{C}ombinatorial \textbf{U}pper \textbf{C}onfidence \textbf{B}ound), an extension of previously proposed algorithm \cite{auer2002finite}. \textbf{CUCB} selects super-actions based on their estimation and variation, i.e., $ \mu_{i,t} + \frac{c\sigma}{\sqrt{T_{i,t}}}$. Next, we are going to prove that \textbf{CUCB} has twice regret bound than \textbf{BLAG} with given round $T$. Similar to \textbf{BLAG}, we have
\vspace{-1.5mm}
\begin{equation}
\small
\begin{split} \mathbb{E}[\overrightarrow{D}\cdot\overrightarrow{\Delta\beta}^t-\alpha\overrightarrow{D}\cdot\overrightarrow{\Delta\beta}^*]\leq 2c\sigma\sum_{i\in\mathcal{S}_t} \frac{1}{\sqrt{T_{i,t}}}+ B^\times\mathcal{P}(\overline{\mathcal{F}_t})
\end{split}	
\end{equation}
\vspace{-3mm}
Under the condition of $\mathcal{F}_t$, $\overrightarrow{D}\cdot\overrightarrow{\beta}_i - 2\frac{c\sigma}{\sqrt{T_{i,t}}} \leq  \mu_{i,t} - \frac{c\sigma}{\sqrt{T_{i,t}}} \leq \overrightarrow{D}\cdot\overrightarrow{\beta}_i$. Thus the minimum reward in second term must be smaller than the actual minimun. Here $c$ and $\mathcal{F}_t$ follow the same definitions as those in \textbf{BLAG}.\\
Now we can obtain the regret bound of \textbf{CUCB} as follows:
\vspace{-2mm}
\begin{displaymath}
\small
\begin{split}
&\mathbb{E}[R_{\textbf{CUCB}}] = \mathbb{E}\left[\overrightarrow{D}\cdot\overrightarrow{\Delta\beta}^t-\alpha\overrightarrow{D}\cdot\overrightarrow{\Delta\beta}^*\right] \leq 2 c\sigma \sum_{t=1}^{T}\sum_{i\in\mathcal{S}_t} \frac{1}{\sqrt{T_{i,t}}} \\& + B^\times\sum_{t=1}\cdot\mathcal{P}(\overline{\mathcal{F}_t})\leq 4c\sigma  M\sqrt{T}+ 1
\end{split}	
\end{displaymath}
Considering that $m$ is a sufficiently large number, and $M=O(m^2)$ while $B^\times=O(m)$, we can see that $\mathbb{E}[R_{\textbf{CUCB}}]=4c\sigma  M\sqrt{T}+ 1\approx 4c\sigma  M\sqrt{T}$ and $\mathbb{E}[R_{\textbf{BLAG}}]=2c\sigma  M\sqrt{T} + 2B^\times \sqrt{T}+1\approx 2c\sigma  M\sqrt{T}$. Comparing to $\mathbb{E}[R_{\textbf{BLAG}}]$, we can see that \textbf{BLAG} has half of the regret bound of \textbf{CUCB} when $T$ is limited. This is because \textbf{CUCB} spends most of its first few rounds on exploration, thus receiving almost random results. Meanwhile, \textbf{BLAG} can limit the exploration procedures by given $\varepsilon_0$, thus having a comparatively stable performance on getting smaller reward.
\vspace{-3mm}
\section{Experimental Results\footnote{ Implementation codes and datasets can be found at https://github.com/EugeneLYC/AAAI18.}}
\subsection{Datasets}
We empirically evaluate the performance of \textbf{BLAG} on one synthetic and three real world datasets\footnote{The three real datasets are adopted from http://snap.stanford.edu/data/index.html}, of which the basic descriptions and statistics are summarized as follows:
\begin{itemize}
	\item\vspace{-2mm} \textbf{B.A. graph} (short for Barabasi Albert graph): This is a synthetic graph that forms with newly added nodes preferentially attached to existing
	 nodes of higher degrees. It well captures the attribute of power-law degree distribution in real social networks. Each graph has two inputs: $n$ (total number of nodes), $p$ (number of edges each new node attaches to existing nodes) with seed nodes fixed. All the B.A. graphs in our experiment are generated using python package `networkx'.
	\item\vspace{-0.5mm} \textbf{Facebook:} This dataset consists of `circles' (or `friends lists') from Facebook. Facebook data was collected from survey participants using this Facebook app. The dataset includes 4,039 node and 88,234 edges.
	\item\vspace{-0.5mm} \textbf{Livejournall:} LiveJournal is a free on-line community with almost 10 million members. This datasets contains 4,847,571 nodes and 68,993,773 edges.
	\item\vspace{-0.5mm} \textbf{Pokec:} As the most popular on-line social network in Slovakia, this dataset contains anonymized data of the whole network, with 1,632,803 nodes and 30,622,564 edges.
\end{itemize}
\vspace{-4.5mm}
\subsection{Information Loss}
We first evaluate the information loss brought by \textbf{BLAG}. In this experiment, we randomly choose 10 connected seeds as sensitive nodes and randomly block 50\% of their neighbors. Here we simulate the learning rounds in one time-slot, thus sensitive nodes are static during the experiment. Each sensitive node transmits signal to its uninformed neighbors in each round, with some random rounds dedicated to sensitive information transmission that is represented by transmitting labeled signal. Let \textit{info} be the cumulative summation of degree of the nodes receiving labels in each round. And we calculate variation of \textit{info} between strategy used and original transmitting per adaptive diffused labeled signal, recording as \textit{info loss} with normalized results. \textit{Info loss} demonstrates to what extent a strategy lose information with same amount of sensitive information being adaptively diffused. To demonstrate the superiority of \textbf{BLAG} in terms of less info loss, we compare it with the two baselines:
\begin{itemize}
	\item \textbf{Monotone Decreasing:} This strategy is originated from \cite{xu2015modeling}, which shows that forwarding information with decreasing probability vs. time limits the cascading size of information. We fit this strategy into our scenario by decreasing transmission probability in a monotone manner with time decreasing amount reduce with time as well.
	\item \textbf{RIPOSTE-like:} This strategy is originated from the previously proposed RIPOSTE\cite{giakkoupis2015privacy}. Typically, RIPOSTE forwards an item with probability slightly larger than a given amount if a user likes the item, or with probability slightly smaller than that amount otherwise. To fit this strategy into our scenario, we randomly choose some rounds and decrease the transmission probability by a static amount.
\end{itemize}
\begin{figure}[!h]\centering
	\vspace{-5mm}\includegraphics[width=0.5\textwidth]{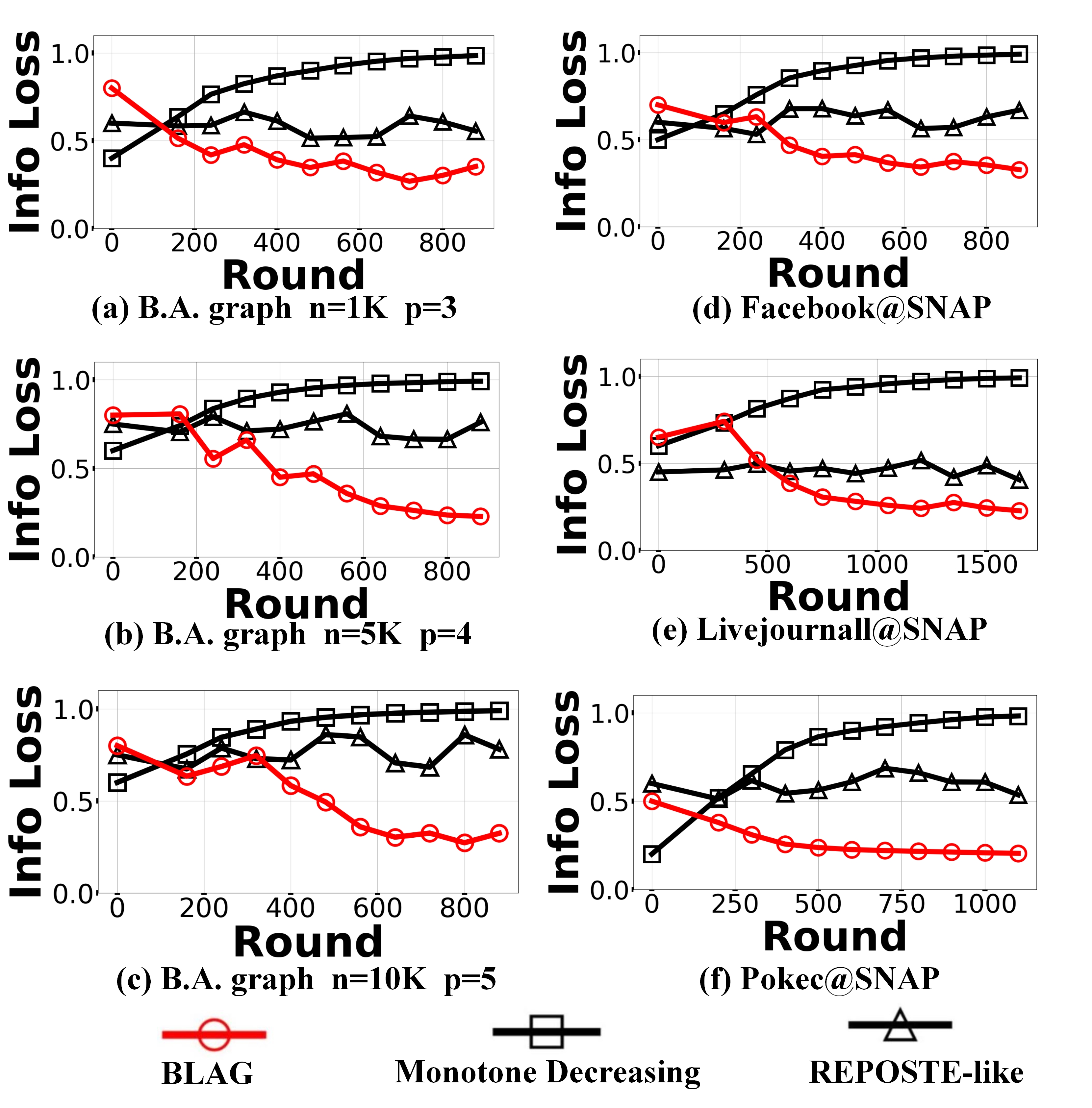}
	\vspace{-7mm}\caption{\small Information loss incurred by different strategies.}	
\vspace{-4mm}\end{figure}
\indent We graphically plot the results of info loss in Fig. 4, where \textbf{BLAG} exhibits the best performance on saving information loss under the same amount of adaptively diffused sensitive information. Due to the noise brought by the learning process, we can see that in some cases depicted by Fig. 4(b)(c)(e), \textbf{BLAG} does not show its effectiveness in the first few rounds, but the trend of \textit{info loss} in \textbf{BLAG} is decreasing. In contrast, RIPOSTE-like strategy has unstable performance. This is caused by the random choice on the rounds. In other words, RIPOSTE-like strategy generally does not consume too much information loss, but meanwhile fails to accurately capture the sensitive information. This result accords with previous assertion that sensitive information is hard to capture. Among all the three strategies, monotone decreasing turns out to perform the worst in the sense that it incurs most information loss in adaptive diffusion of each piece of sensitive information.\\
\indent Furthermore, Fig. 4(a)(b)(c) reveal with the increase of network scale and complexity, that is, more nodes (increasing of $m$) and larger density (increasing of $p$), \textbf{BLAG} goes through a longer sequence of rounds with unpredictable \textit{info loss}. This is caused by increasing of $m$, thus leading to larger size of $ASG$. That results in randomness in combination adopted from each round and consumption of more rounds to get close to the optimal action.\\
\indent From Fig. 4(a)-(f) we can roughly calculate the least \textit{info loss} saved by \textbf{BLAG} and other two algorithms. The least amount is depicted in Fig. 4(e), where we get 0.25 normalized info loss from \textbf{BLAG} and 0.4 from REPOSTE-like. Their difference is about 40\%.
\vspace{-3mm}
\subsection{Regret in Learning Process}
Now we proceed to evaluate the regret of \textbf{BLAG} and \textbf{CUCB} with limited rounds and large base-action set. Particularly, as it is time consuming to calculate the regret bound, we calculate the cumulative reward of each round instead. Note that the two evaluations are equivalent since the regret can be deduced from the reward. First we extract $m$ nodes from a B.A. graph ($n=10K,p=5$) as \emph{target nodes} and record their degrees. Then we generate $2\binom{m}{2}$ base-actions by randomly allocating element values under conditions discussed in previous sections. Thus $2\binom{m}{2}$ is the size of base-action set. And we set $\varepsilon_0=1$ for the initialization of global $\varepsilon-greedy$ process. The comparison results are summarized in Table 3\\
\vspace{-3mm}
\begin{table}[!h]\footnotesize
	\centering
	\begin{tabular} {p{1.3cm}<{\centering}p{1.8cm}<{\centering}p{1.8cm}<{\centering}p{1.8cm}<{\centering}}
		\hline
		Rounds &Base-actions  & \textbf{CUCB} & \textbf{BLAG}\\ \hline
\multirow{3}{*}{$1000$}
&200    & -0.6e-2  & -6.7e-2\\
&5K    & -0.6e-3  & -32.0e-3\\
&20K    & -0.4e-3  & -4.8e-3\\
&2M    & -0.2e-5  & -12.3e-4\\ \hline
\multirow{3}{*}{$3000$}
&200    & -1.4e-2  & -8.4e-2\\
&5K    & -0.8e-3  & -22.3e-3\\
&20K    & -0.5e-3  & -109.9e-3\\
&2M    & -1.6e-4  & -14.8e-4\\
		\hline
	\end{tabular}
	\caption{\small Comparison of cumulative reward between \textbf{CUCB} and \textbf{BLAG} within limited rounds. The results are normalized.}
\end{table}
\indent From Table 3 we observe a superior learning performance of the global $\varepsilon-greedy$ process adopted in \textbf{BLAG} to \textbf{CUCB}. This is because in the first few rounds, $\varepsilon-greedy$ can allocate rounds for exploration and exploitation by the given $\varepsilon_0$ while \textbf{CUCB} allocates rounds by the observed reward and variation of each base-action. Due to the limited rounds, most of the base-actions in $ASG$ are chosen with few times. That means, \textbf{CUCB} spends most of its first few rounds on exploration, thus leading to unstable (bad) performance. By comparison, $\varepsilon-greedy$ benefits from the current estimation and artificially limited exploration probability. In real life, the initiated $\varepsilon_0$ can be determined by algorithm excuters to deceide whether stick to the current estimation or slightly tolerate more explorations. We can see from Table 3 that given same rounds, \textbf{BLAG} gets reward at least 10 times than \textbf{CUCB} (remember in our scenario, the smaller reward, the better).\\
\indent Furthermore, with number of base-actions increase from 200 to 2M, \textbf{CUCB} has an obvious deteriorating performance, increasing from -0.6e-2 to -0.2e-5 and from -1.4e-2 to -1.6e-4 when learning rounds are 1,000 and 3,000, respectively. In contrast, \textbf{BLAG} has a stable performance on getting reward. We can see that when given 3,000 rounds, cumulative reward from \textbf{BLAG} decrease from -22.3e-3 to -109.9e-3 when base-actions increase from 5K to 20K. This is also attributed to the increased randomness in \textbf{CUCB} and \textbf{BLAG} sticking to the current estimation controlled by given $\varepsilon$. And we can slightly make the conclusion that with greater extent on large base-action set and limited rounds, \textbf{BLAG} will show even more superior performance than \textbf{CUCB}.
\vspace{-3mm}
\subsection{Cascading Scale}
Last but not least, we also take a look into the effectiveness of adaptive diffusion on limiting cascading of sensitive information. In doing so, in each experiment, we first randomly select a source node to be the sensitive node. We initialize the transmission probability on each edge connecting a sensitive node and non-sensitive node be $5e-5$, which means the sensitive nodes have $5e-5$ probability to transfer its non-sensitive nodes to be sensitive ones. At each time slot, we sort the degree of neighbors and let transmission probability to those first half nodes be $0$ and to those last half be  $1e-4$. In that case, we adaptively diffuse the sensitive information but maintain the overall probabilities, which accords to the main idea of adaptive diffusion. Since the learning process is not the focus in this experiment, we let total number of round in one time-slot be 1. And we calculate the rate of sensitive nodes divided by total number of nodes in the network.
\begin{figure}[!h]\centering
	\includegraphics[width=0.48\textwidth]{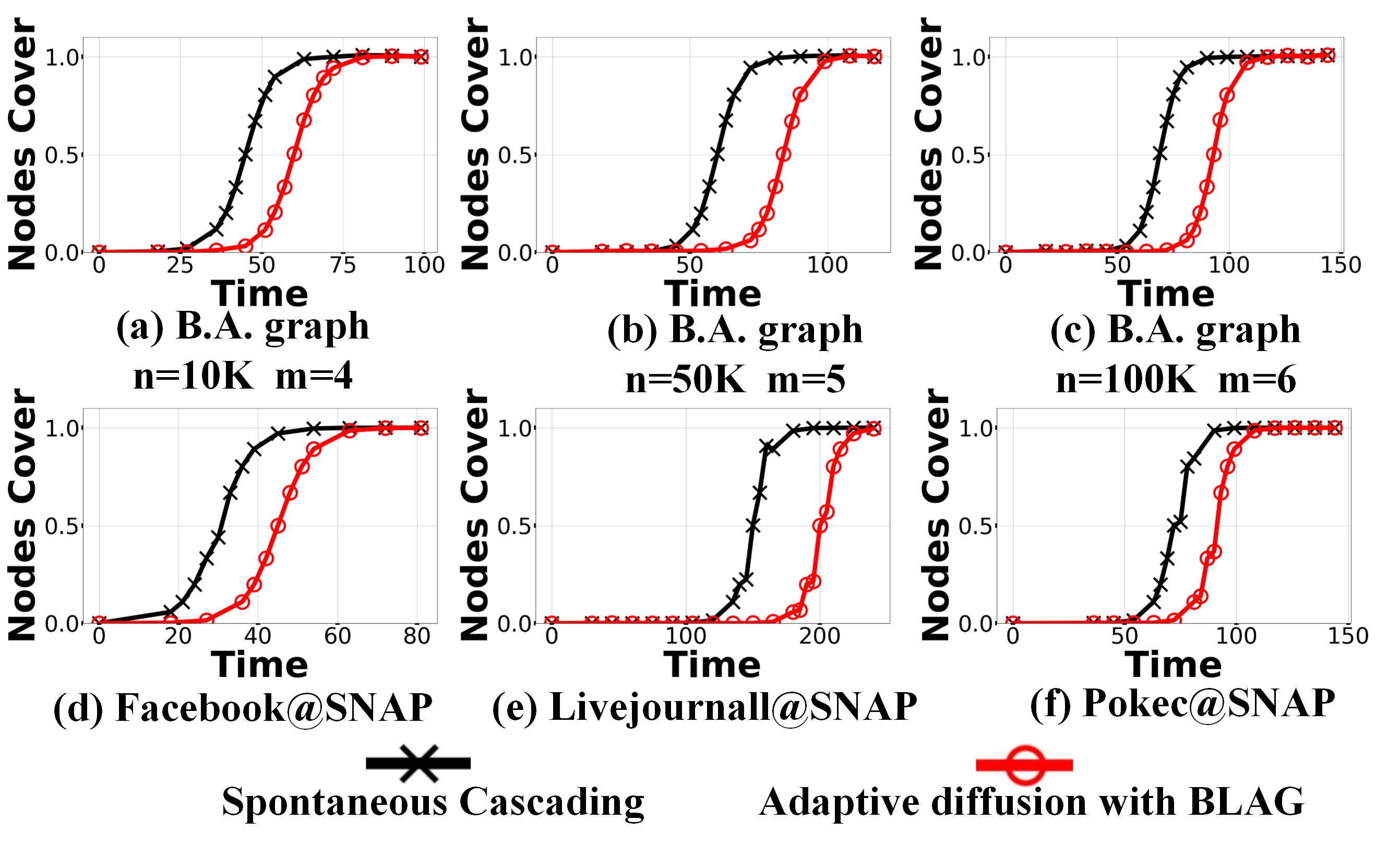}
	\vspace{-7.5mm}\caption{\small Cascading size under adaptive diffusion of \textbf{BLAG} and spontaneous spreading.}
\vspace{-4mm}\end{figure}
Fig. 4 plots the cascading scale of both \textbf{BLAG} and spontaneous spreading under B.A. graphs (Fig. 4(a)-(c)) and three real datasets (Fig. 4(d)-(f)). A common observation from Fig. 4(a)-(f)  is that the cascading behavior undergoes a transition, i.e., the cascading scale increases explosively after a certain time threshold. Adaptive diffusion has strong effect on postponing such threshold. Considering of the time-sensitive attribute of sensitive information, if the threshold is postponed after the timeliness of sensitive information, no sensitive information will be widely spread.
\vspace{-5mm}
\section{Related Works}
Back in 2003, \cite{kempe2003maximizing} analyzed two classic diffusion models: LTM and ICM. Different works have been done on information diffusion.  \cite{cheng2014can}\cite{prakash2012threshold}\cite{venkatramanan2012co} discuss how different behavior in network may affect the information diffusion. \cite{iribarren2009impact} and \cite{karsai2011small} discuss human activities patterns that would slow down diffusion. \cite{myers2012information} and \cite{wu2011says} convey investigation on how roles of users influence information diffusion.

Our work is also connected to combinatorial multi-arm bandit problem. In our combinatorial bandit setting, feedback is observed on each played arm individually during each round of play. Other types of feedback \cite{audibert2011minimax}\cite{mannor2011bandits} include observing the outcomes of all arms, and only observeing the final reward but no outcome of any individual arm.  \cite{anantharam1987asymptotically}\cite{caro2007dynamic} consider simultaneous plays of arms. \cite{auer2002using}\cite{hanawal2015cheap} discuss the case where only the cumulative reward of the combination can be observed in a single round.
\vspace{-6mm}
\section{Conclusion}
\indent In this paper, we propose an adaptive diffusion strategy in social network to protect sensitive information while decreasing information loss. We design \textbf{BLAG} algorithm to tackle the formed constrained combinatorial Multi-Armed Bandit problem. We show that our algorithm reduces time-limited regret bound by half compared to previous proposed \textbf{CUCB} algorithms.

\section{Appendix}
\textbf{Proof of Theorem 1:}
Let vector $\overrightarrow{y}^*$ be
\begin{displaymath}
\overrightarrow{y}^*(j) \triangleq \left\{
\begin{aligned}
1&, \quad1\leq j \leq B_0 \\
0&, \quad B_0 < j \leq m
\end{aligned}\right.
\end{displaymath}
thus,
\begin{displaymath}
\sum_{j=1}^{m}(\overrightarrow{\beta}(j)  - \overrightarrow{y}^*(j) ) = 0
\end{displaymath}
If $1\leq j \leq B_0$, $\overrightarrow{\beta}(j)  - \overrightarrow{y}^*(j)\leq 0$. Similarly, if $B_0< j \leq m$ holds, $\overrightarrow{\beta}(j)  - \overrightarrow{y}^*(j)\geq 0$.
thus,
\begin{displaymath}
\begin{aligned}
\small
&\overrightarrow{D}^T\overrightarrow{\beta}-  \overrightarrow{D}^T\overrightarrow{y}^* \\
&=  \sum_{j=1}^{B_0}\overrightarrow{D}(j)(\overrightarrow{\beta}(j)  - \overrightarrow{y}^*(j)) \\& + \sum_{j=B_0+1}^{m}\overrightarrow{D}(j)(\overrightarrow{\beta}(j)  - \overrightarrow{y}^*(j)) \\
&\geq \overrightarrow{D}(B_0) \sum_{j=1}^{B_0}(\overrightarrow{\beta}(j)  - \overrightarrow{y}^*(j)) \\&+  \overrightarrow{D}(B_0+1)\sum_{j=B_0+1}^{m}(\overrightarrow{\beta}(j)  - \overrightarrow{y}^*(j)) \\
&\geq \overrightarrow{D}(B_0+1) -\overrightarrow{D}(B_0)]\sum_{j=B_0+1}^{m}(\overrightarrow{\beta}(j)  - \overrightarrow{y}^*(j))  \geq 0
\end{aligned}
\end{displaymath}
Finally,we can get
\begin{displaymath}
\small
\overrightarrow{D}^T\overrightarrow{\beta}- \overrightarrow{D}^T\overrightarrow{\beta}_0 \geq \overrightarrow{D}^T\overrightarrow{y}^*-\overrightarrow{D}^T\overrightarrow{\beta}_0 =  B^*
\end{displaymath}$\hfill\blacksquare$\\
\textbf{Proof of Lemma 1:}
Similar to Theorem 1, let
\begin{displaymath}
\small
\overrightarrow{z}^*(j) \triangleq \left\{
\begin{aligned}
1&, \quad1\leq j \leq m - B_0 \\
0&, \quad m - B_0 < j \leq m
\end{aligned}\right.
\end{displaymath}
and
\begin{displaymath}
\small
\forall \overrightarrow{\beta}: \sum_{j=1}^{m}\beta(j) = B_0, \quad \overrightarrow{D}^T\overrightarrow{\beta} \geq \sum_{j=m-B_0 +1}^{m} \overrightarrow{D}(j)
\end{displaymath}
we can get
\begin{displaymath}
\small
\overrightarrow{D}^T\overrightarrow{\beta_1} - \overrightarrow{D}^T\overrightarrow{\beta_2}  \leq \sum_{j=m-B_0+1}^{m} \overrightarrow{D}(j) -\sum_{j=1}^{B_0} \overrightarrow{D}(j)  = B^\times
\end{displaymath} $\hfill\blacksquare$\\
\textbf{Deduction of equation (1):}\\
We can write expression under event $\mathcal{F}_t$ in \textbf{BLAG} as
\begin{displaymath}
\small
\begin{aligned}
&\mathop{\mathbb{E}}_{\mathcal{F}_t}[\overrightarrow{D}^T\overrightarrow{\Delta\beta}_{et}^t-\alpha\overrightarrow{D}^T\overrightarrow{\Delta\beta}^*]\\
& \leq
\mathop{\mathbb{E}}_{\mathcal{F}_t}\left[\overrightarrow{D}^T\overrightarrow{\Delta\beta}_{et}^t-\sum_{i\in\mathcal{S}_t } \mu_{i,t}\right]  + \mathop{\mathbb{E}}_{\mathcal{F}_t}\left[ \sum_{i\in\mathcal{S}_t }\mu_{i,t}\right] -\alpha\overrightarrow{D}^T\overrightarrow{\Delta\beta}^*\\
&\leq  \mathop{\mathbb{E}}_{\mathcal{F}_t}\left[\sum_{i\in\mathcal{S}_t }\left|\overrightarrow{D}^T\overrightarrow{\beta}_i - \mu_{i,t}\right|\right]  + \alpha\left(\mathop{\mathbb{E}}_{\mathcal{F}_t}\left[ \sum_{i\in\mathcal{S}^*_t }\mu_{i,t}\right] -\overrightarrow{D}^T\overrightarrow{\Delta\beta}^*\right)\\
&\leq c\sigma\sum_{i\in\mathcal{S}_t} \frac{1}{\sqrt{T_{i,t}}}
\end{aligned}
\end{displaymath}
Thus, we can get the expectation in overall situation
\begin{displaymath}
\small
\begin{aligned}
&\mathop{\mathbb{E}}[\overrightarrow{D}^T\overrightarrow{\Delta\beta}_{et}^t-\alpha\overrightarrow{D}^T\overrightarrow{\Delta\beta}^*]\leq c\sigma\sum_{i\in\mathcal{S}_t} \frac{1}{\sqrt{T_{i,t}}}+ B^\times\mathcal{P}(\overline{\mathcal{F}_t})
\end{aligned}
\end{displaymath}$\hfill\blacksquare$\\
\textbf{Deduction of equation (2):}
\begin{displaymath}
\small
\begin{aligned}
\mathbb{E}\left[\sum_{t=1}\cdot\overrightarrow{D}\cdot\overrightarrow{\Delta\beta}_{et}^t-\alpha\overrightarrow{D}\cdot\overrightarrow{\Delta\beta}^*\right]
\leq c\sigma \sum_{t=1}^{T}\sum_{i\in\mathcal{S}_t} \frac{1}{\sqrt{T_{i,t}}}  + B^\times\sum_{t=1}\cdot\mathcal{P}(\overline{\mathcal{F}_t})\\
\leq c\sigma  \sum_{i\in\mathcal{V}}\sum_{l=1}^{T_{i,T}-1} \frac{1}{\sqrt{l}} + B^\times\sum_{t=1}\cdot\mathcal{P}(\overline{\mathcal{F}_t})
\leq 2c\sigma  M\sqrt{T}+ B^\times\sum_{t=1}\cdot\mathcal{P}(\overline{\mathcal{F}_t})
\end{aligned}
\end{displaymath}
The third equality exchanges the summation order, and notice when $i\in\mathcal{S}_t$, $T_{i,t+1}=T_{i,t}+1$. The fourth equality holds because $T_{i,T}-1 \leq T$. $M$ is the initial size of $ASG$ where $M \propto \binom{m}{2}\hfill\blacksquare$.\\
\textbf{Deduction of equation (3):}\\
We can write expression under event $\mathcal{F}_t$ in \textbf{CUCB} as
\begin{displaymath}
\small
\begin{split}
&\mathop{\mathbb{E}}_{\mathcal{F}_t}[\overrightarrow{D}^T\overrightarrow{\Delta\beta}^t-\alpha\overrightarrow{D}^T\overrightarrow{\Delta\beta}^*] \leq
\mathop{\mathbb{E}}_{\mathcal{F}_t}\left[\overrightarrow{D}^T\overrightarrow{\Delta\beta}^t - \sum_{i\in\mathcal{S}_t }\left( \mu_{i,t} - \frac{c\sigma}{\sqrt{T_{i,t}}}\right)\right] \\&+ \mathop{\mathbb{E}}_{\mathcal{F}_t}\left[ \sum_{i\in\mathcal{S}_t }\left( \mu_{i,t} - \frac{c\sigma}{\sqrt{T_{i,t}}}\right)\right] -\alpha\overrightarrow{D}^T\overrightarrow{\Delta\beta}^*\\&\leq  \mathop{\mathbb{E}}_{\mathcal{F}_t}\left[\sum_{i\in\mathcal{S}_t }\left|\overrightarrow{D}^T\overrightarrow{\beta}_i -  \left( \mu_{i,t} - \frac{c\sigma}{\sqrt{T_{i,t}}}\right)\right|\right] \\&+ \alpha\left(\mathop{\mathbb{E}}_{\mathcal{F}_t}\left[ \sum_{i\in\mathcal{S}^*_t }\left( \mu_{i,t} - \frac{c\sigma}{\sqrt{T_{i,t}}}\right)\right] -\overrightarrow{D}^T\overrightarrow{\Delta\beta}^* \right)\leq 2c\sigma\sum_{i\in\mathcal{S}_t} \frac{1}{\sqrt{T_{i,t}}}
\end{split}	
\end{displaymath}
Thus, we can get the expectation in overall situation
\begin{displaymath}
\small
\begin{split}
&\mathbb{E}[\overrightarrow{D}^T\overrightarrow{\Delta\beta}^t-\alpha\overrightarrow{D}^T\overrightarrow{\Delta\beta}^*]\leq 2c\sigma\sum_{i\in\mathcal{S}_t} \frac{1}{\sqrt{T_{i,t}}}+ B^\times\mathcal{P}(\overline{\mathcal{F}_t})
\end{split}	
\end{displaymath}$\hfill\blacksquare$

\bibliographystyle{aaai}
\bibliography{cited}
\end{document}